\documentclass[entropy,article,submit,moreauthors,pdftex,10pt,a4paper]{mdpi} 

\firstpage{1} 
\articlenumber{x}
\doinum{10.3390/------}
\pubvolume{xx}
\pubyear{2017}
\copyrightyear{2017}
\externaleditor{Academic Editors: Andrea Puglisi, Alessandro Sarracino and Angelo Vulpiani}
\history{Received: date; Accepted: date; Published: date}

\preto{\abstractkeywords}{\nolinenumbers}

\Title{A Chain, a Bath, a Sink and a Wall
}

\Author{S. Iubini $^{1,2}$*, S. Lepri $^{3,2}$, R. Livi $^{1,2,3}$, G.-L. Oppo  $^{4}$ and A. Politi $^{5}$}

\AuthorNames{Stefano Iubini, Stefano Lepri, Roberto Livi, Gian-Luca Oppo and Antonio Politi}

\address{%
$^{1}$ \quad Dipartimento di Fisica e Astronomia, Universit\`a di Firenze, 
via G. Sansone 1 I-50019, Sesto Fiorentino, Italy\\
$^{2}$ \quad Istituto Nazionale di Fisica Nucleare, Sezione di Firenze, 
via G. Sansone 1 I-50019, Sesto Fiorentino, Italy\\
$^{3}$ \quad Consiglio Nazionale delle Ricerche, 
Istituto dei Sistemi Complessi, via Madonna del Piano 10, I-50019 Sesto Fiorentino, Italy\\
$^{4}$\quad SUPA and Department of Physics, University of Strathclyde, Glasgow G4 ONG, Scotland, United Kingdom \\
$^{5}$ \quad  Institute for Complex Systems and Mathematical Biology \& SUPA
University of Aberdeen, Aberdeen AB24 3UE, United Kingdom \\
}

\corres{Correspondence: stefano.iubini@unifi.it}

\abstract{We investigate out-of-equilibrium stationary processes emerging in a 
Discrete Nonlinear Schr\"odinger chain in contact with a heat reservoir (\textit{a bath}) 
at temperature $T_L$ and a pure dissipator (\textit{a sink}) acting on opposite edges. 
We observe two different regimes. For small heat-bath temperatures $T_L$ and chemical-potentials, 
temperature profiles across the chain display a non-monotonous shape, remain remarkably 
smooth and even enter the region of \textit{negative absolute temperatures}.
For larger temperatures $T_L$, the transport of energy is strongly inhibited by the spontaneous
emergence of \textit{discrete breathers}, which act as a thermal \textit{wall}. A strongly
intermittent energy flux is also observed, due to the irregular birth and death events of
the breathers. The corresponding statistics exhibits the typical signature of rare events 
of processes with large deviations. In particular, the breather lifetime is found to be ruled by a 
stretched-exponential law. }



\begin{document}


\section{Introduction}

The study of nonequilibrium thermodynamics of systems composed of a relatively small 
number of particles is motivated by the need of a deeper theoretical understanding
of statistical laws leading to the possibility of manipulating small-scale 
systems like biomolecules, colloids or nano-devices. In this framework, statistical
fluctuations and size effects play a major role and cannot be ignored as it is customary 
to do in their macroscopic counterparts.

Arrays of coupled classical oscillators are representative models of 
such systems and have been studied intensively in this context \cite{LLP03,DHARREV,Basile08}.
In particular, the Discrete Nonlinear Schr\"odinger (DNLS) equation has been 
widely investigated in various domains of physics as a prototype 
model for the propagation of nonlinear excitations \cite{Eilbeck1985,Eilbeck2003,Kevrekidis}. 
In fact, it provides an effective description of electronic transport in 
biomolecules \cite{Scott2003} as well as of nonlinear waves propagation in a  
layered photonic or phononic systems \cite{Kosevich02,Hennig99}. More recently, a 
renewed interest for this multipurpose equation emerged in the physics of 
gases of ultra-cold atoms trapped in optical lattices
(e.g., see Ref. \cite{Franzosi2011} and references therein for a recent survey).
Since the seminal paper by Rasmussen et al. \cite{Rasmussen2000}
it was realized that the presence in the DNLS of intrinsically localized 
solutions, named {\it discrete breathers} (DB) (see e.g. \cite{Sievers1988,MacKay1994,Flach2008}),
could be associated to negative absolute temperature states. 
In a series of important papers, Rumpf provided entropy-based arguments to 
describe asymptotic states above a modulational instability line in the DNLS 
\cite{Rumpf2004,Rumpf2008,Rumpf2009,Rumpf2007}.
It has been later found that above this line, negative temperature states can form 
spontaneously via the dynamics of the DNLS. They persist over extremely long time scales, 
that might grow exponentially with the system size as a result of an effective
mechanism of ergodicity--breaking \cite{Iubini2013}. It has been recognized that
most of these peculiar features of the DNLS can be traced back to the
properties of its Hamiltonian that admits two first integrals of motion: the total
energy $H$ and the total number $A$ of particles (also termed as the total norm). 

A related question is how the structure of the Hamiltonian influences non-equilibrium properties 
when the system can exchange energy and/or mass with the environment. 
In a series of papers~\cite{Livi2006, Franzosi2007,Iubini2013} it has been found that,
when pure dissipators act at both edges of a DNLS chain (a case sometimes called 
boundary cooling \cite{Tsironis1996,Piazza2001,Piazza2003,Reigada2003}), 
the typical final state consists in an isolated static breather embedded in an almost empty background. 
The breather collects a sensible fraction of the initial energy and it is essentially
decoupled from the rest of the chain. The spontaneous creation of localized energy spots 
out of fluctuations has further consequences on the relaxation to equipartition, 
since the interaction with the remaining part of the chain can become exponentially 
weak~\cite{DeRoeck2013,Cuneo2016}.
A similar phenomenology occurs after a quench from high to low temperatures 
in oscillator lattices \cite{Oikonomou2014}. Also, boundary driving by external 
forces may induce non-linear localization \cite{Geniet2002,Maniadis2006,Johansson2009}.

When, instead, the chain is put in contact with thermal baths at its edges, stationary states 
characterized by a gradient of temperature and chemical potential emerge~\cite{Iubini2012}.
The transport of mass and energy is typically normal (diffusive) and can be described in terms
of the Onsager formalism. However, peculiar features such as non-monotonous temperature 
profiles~\cite{Iubini2013a}, or persistent currents~\cite{Borlenghi2015} are found, as well as
a signature of anomalous transport in the low-temperature regime~\cite{Kulkarni2015,Mendl2015}.

In this paper we consider a setup where one edge is in contact with a heat reservoir at temperature $T_L$
and chemical potential $\mu_L$,
while the other interacts with a \textit{pure dissipator}, i.e. a mass sink.
The original motivation for studying this configuration was to better understand the 
role of DBs in thermodynamic conditions. At variance with standard setups \cite{DHARREV,LLP03},
this is conceptually closer to a semi-infinite array in contact with a single reservoir.
In fact, on the pure-dissipator side, energy can only flow out of the system. 

\begin{figure}[ht]
\begin{center}
\includegraphics[width=9 cm,clip]{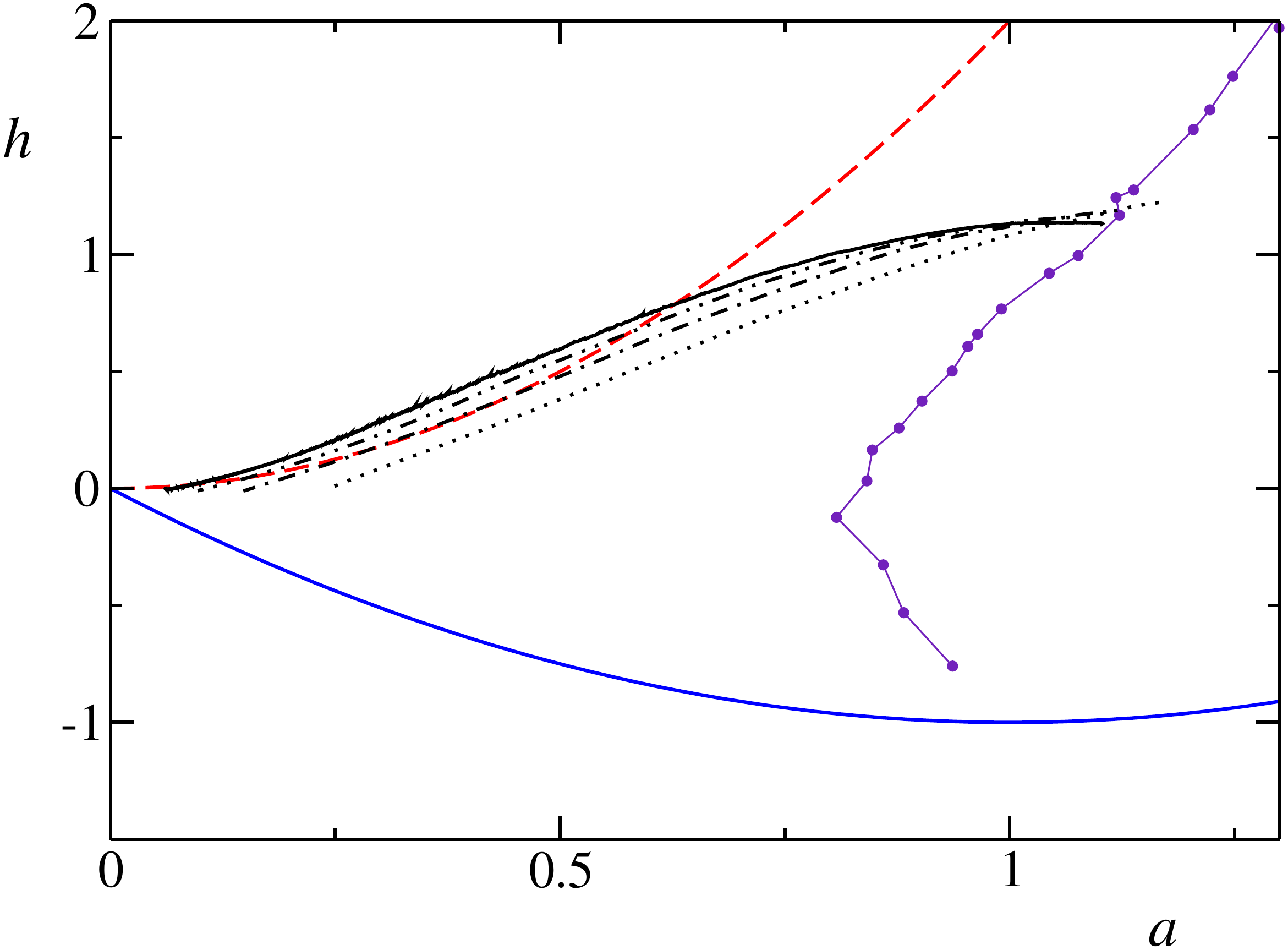}
\end{center}
\caption{Phase diagram of the DNLS equation in the ($a-h$) plane of, respectively, energy and mass 
densities. The positive-temperature region extends between the ground state $\beta=+\infty$ line 
(solid blue lower curve) and the $\beta=0$ isothermal (red dashed curve). Purple circles 
show the $\mu = 0$ line, that has been determined numerically through equilibrium simulations 
(see Ref. ~\cite{Iubini2013a}).
Black curves refer to nonequilibrium profiles obtained by employing a heat bath with parameters 
$T_L=3$ and $\mu_L=0$ and a pure dissipator located
respectively at the left and right edges of the chain. 
Dotted, dot-dashed, dot-dot-dashed and solid curves refer to chain sizes $N=511$, 1023, 2047 and 4095. 
Upon increasing $N$, these above profiles tend to enter the negative temperature region,
cfr also Fig. \ref{fig14}. 
Simulations are performed by evolving the DNLS chain over $10^7$ time units after a 
transient of $4\times10^7$ units. For the system size $N=4095$, a further average over 10 independent 
trajectories is performed.
}
\label{fig-a-h}
\end{figure}

A non-equilibrium steady state can be conveniently represented as a path 
in the $(a-h)$-parameter space, where $a(x)$ is the mass density, $h(x)$ the
energy density, and $x\in[0,1]$ the rescaled position along the chain  
(see Fig.~\ref{fig-a-h} for a few sampled paths). 
Making use of suitable microcanonical definitions~\cite{Iubini2012}, these paths can be 
converted into temperature ($T$) and chemical potential ($\mu$) profiles. 
The presence of a pure dissipator in $x=1$ forces the corresponding path to terminate close 
to the point $(a=0,h=0)$, which is singular both in $T$ and $\mu$.
Therefore, slight deviations may easily lead to crossing the $\beta=0$ line, 
where $\beta$ is the inverse temperature. This is indeed the typical scenario observed for
small $T_L$, when $\beta$ smoothly changes sign twice before approaching the dissipator.
The size of the negative-temperature region increases with $T_L$.
For high temperatures, a different stationary regime is found, characterized
by strong fluctuations of mass and energy flux. In fact, upon increasing $T_L$,
the negative-temperature region first extends up to the dissipator edge and then it progressively shrinks 
in favour of a positive-temperature region (on the other side of the chain).
In this regime, the dynamics is controlled by the spontaneous formation (birth) and disappearance
(death) of discrete breathers.

In Section~\ref{sec1} we introduce the model and briefly recall the main observables.
Section~\ref{sec2} is devoted to a detailed characterization of the low-temperature phase, while
the strongly non-equilibrium phase observed for large $T_L$ is discussed in Section \ref{sec3}). 
This is followed by the analysis of the statistical properties of the birth/death process of
large-amplitude DBs, illustrated in Section~\ref{sec4}.
Finally, in Section~\ref{sec5} we summarize the main results and
comment about possible relationships with similar
phenomena previously reported in the literature. 

\section{Model and observables}
\label{sec1}
We consider a DNLS chain of size $N$ and with open boundary conditions, whose bulk dynamics is ruled 
by the equation 
\begin{equation}
\label{eqmot}
i \dot{z}_n=-2|z_n|^2z_n -z_{n-1}- z_{n+1}
\end{equation}
where ($n=1,\ldots,N$), $z_n = (p_n + i q_n)/\sqrt{2}$ are complex variables, with $q_n$ and $p_n$ 
being standard conjugate canonical variables. The  quantity $ a_n = |z_n(t)|^2$ 
can be interpreted as the {\sl number of particles}, or, equivalently, the  {\sl mass} in
the lattice site $n$ at time $t$.
Upon identifying the set of canonical variables $z_n$ and $i z_n^*$,
Eq.~(\ref{eqmot}) 
can be read as the equation of motion generated by the Hamiltonian functional 
\begin{equation}
 H= \sum_{n=1}^N \left( |z_n|^4+z_n^*z_{n+1}+z_nz_{n+1}^* \right) \quad 
\label {Hz}
 \end{equation}
 through the 
Hamilton equations $\dot{z}_n=-\partial{H}/\partial{(iz_n^*)}$. 
We are dealing
with a dimensionless version of the DNLS equation:  the nonlinear coupling
constant and the hopping parameters, that usually are indicated explicitly  in the Hamiltonian
(\ref{Hz}), have been set equal to unity. Accordingly, also the time variable $t$ is expressed
in arbitrary adimensional units. Without loss of generality, this formulation
has the advantage of simplifying numerical simulations where we can easily check that 
$H$ and the total mass $A = \sum_n |z_n|^2(t)$ are conserved dynamical quantities.
 
The first site of the chain ($n=1$) is in contact with a reservoir at temperature $T_L$ and chemical 
potential $\mu_L$. 
This is ensured by implementing the non-conservative Monte-Carlo dynamics described in Ref.~\cite{Iubini2012}.  
The opposite site ($n=N$) interacts with a pure \textit{stochastic dissipator}: the variable $z_N$ is set equal 
to zero at random times, whose separations are independent and identically distributed variables uniformly distributed within the interval 
$[t_{min},t_{max}]$.  On average, this corresponds to simulating a dissipation process with decay 
rate $\gamma \sim {\bar t}^{-1}$, where ${\bar t} = (t_{max} + t _{min})/2$. Notice that
different prescriptions, such as for example a Poissonian distribution of times with average $\bar t$,
or a constant pace equal to $\bar t$, do not introduce any relevant modification in the dynamical and
statistical properties of the model (\ref{eqmot}).
Finally, the Hamiltonian dynamics between successive interactions with the thermostats
has been generated by implementing a symplectic, 4th-order Yoshida algorithm \cite{Yoshida1990}.
We have verified that a time step $\Delta t = 2\times 10^{-2}$ suffices to ensure suitable accuracy.

Throughout the paper we deal with measurements of local temperature and chemical
potential. Since the Hamiltonian is non separable, it is necessary to make use of the microcanonical
definition provided in~\cite{Rugh1997}. The general expressions are nonlocal and rather involved;
we refer to \cite{Iubini2012} for details and the related bibliography.  

In what follows, we consider a situation where all parameters, other than $T_L$, are kept
fixed. 
In particular, we have chosen $\mu_L=0$ and  
$\bar t =  3 \times 10^{-2}$, with  $t_{max}$ and $t_{min}$ of order $10^{-2}$. 
We have verified that the results obtained for this choice of the
parameter values are general. A more detailed account of the dependence of the results on the
thermostat properties will be reported elsewhere.

Finally, we recall the observables that are typically used to characterize a steady-state out of 
equilibrium: the mass flux 
\begin{equation}
j_a= 2 \langle \mathrm{Im}(z_n^*z_{n-1}) \rangle \; ,
\end{equation}
and the energy flux 
\begin{equation}
j_h= 2 \langle \mathrm{Re}(\dot z_nz_{n-1}^*) \rangle \; .
\end{equation}

\section{Low--temperature regime: coupled transport and negative temperatures}
\label{sec2}

\begin{figure}[ht]
\begin{center}
\includegraphics[width=7.5 cm,clip]{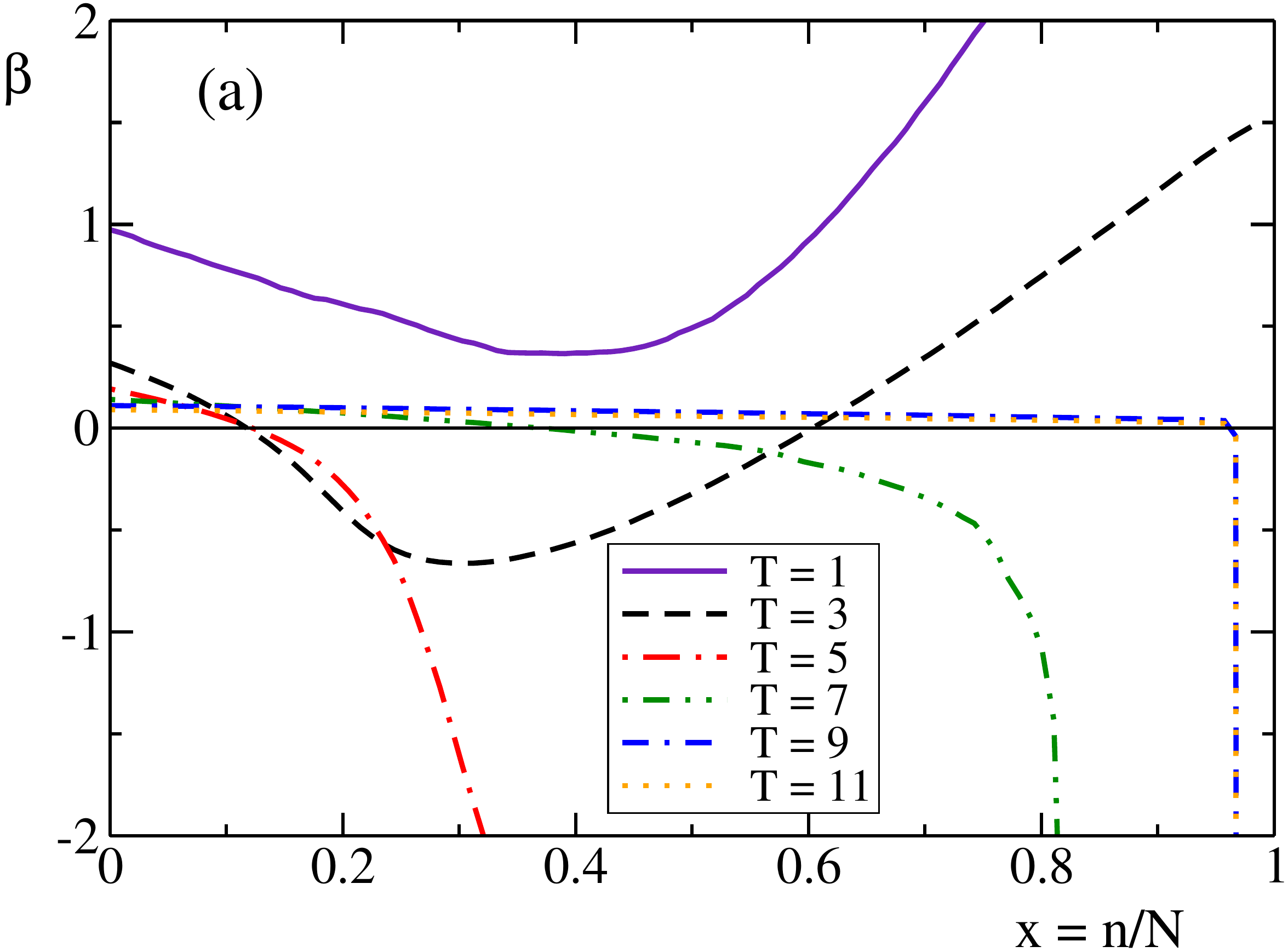}
\includegraphics[width=7.5 cm,clip]{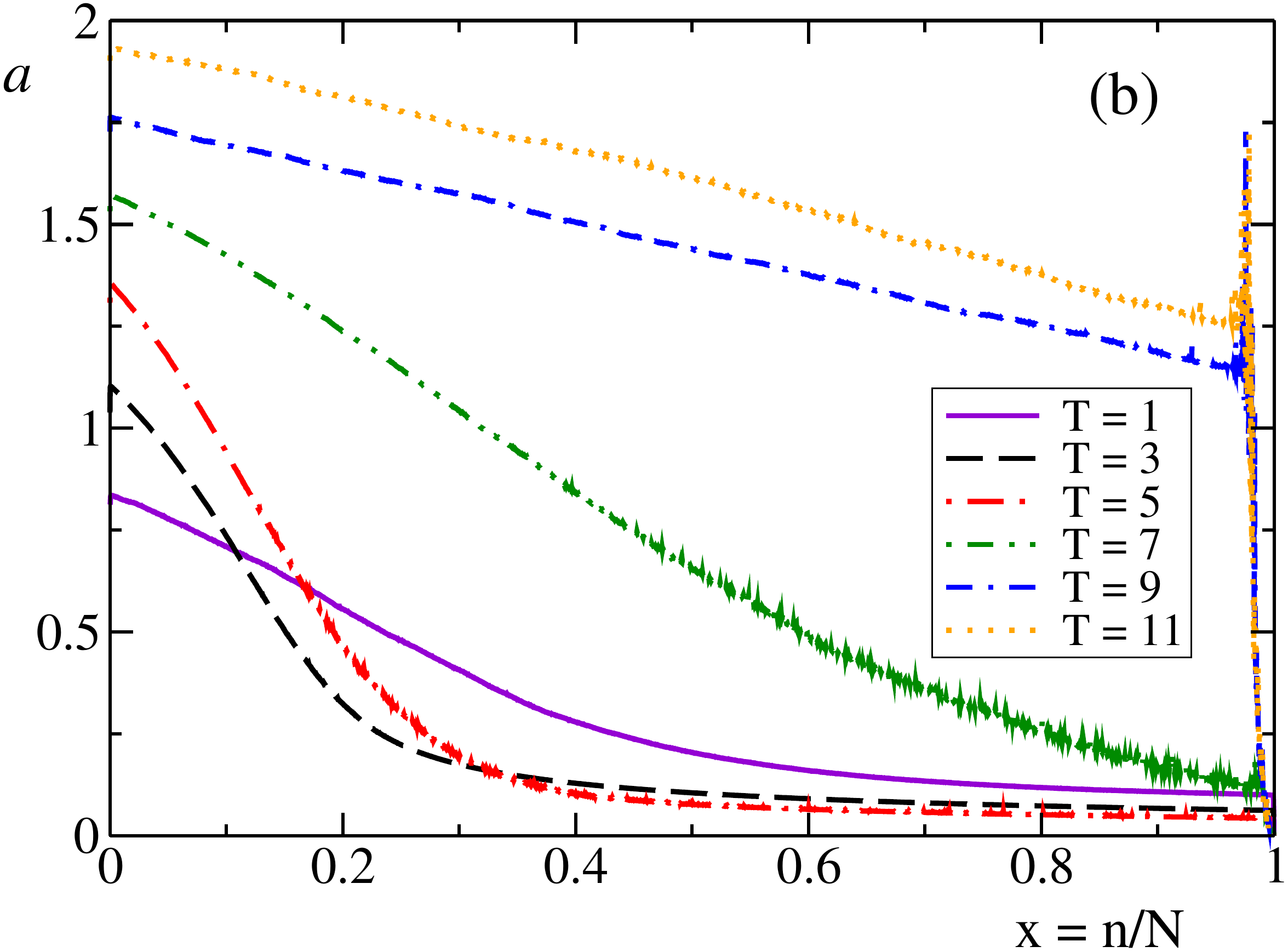}
\end{center} 
\caption{Average profiles of the inverse temperature $\beta(x)$ (panel (a)) 
and mass-density $a(x)$ (panel (b)) for a DNLS chain with $N=4095$ 
and different temperatures $T_L$ of the reservoir acting at the left edge, where $\mu=0$.
The profile $\beta(x)$ is computed making use of the microcanonical definition of temperature.
Simulations are performed evolving the DNLS chain over $10^7$ time units after a transient of 
$4\times10^7$ units. In order to obtain a reasonable smoothing of these time-averaged profiles
we have further averaged each of them over 10 independent trajectories.
}
\label{fig15}
\end{figure}

In the left panel of Fig.~\ref{fig15} we report the average profile of the inverse temperature 
$\beta(x)$ as a function of the rescaled site position $x=n/N$, for different values of the
temperature of the thermostat. A first ``anomaly" is already noticeable for relatively small $T_L$:
the profile is non monotonous (see for example the curve for $T_L=1$). This feature is frequently 
encountered when a second quantity, besides energy, is transported~\cite{Iacobucci2011,Iubini2012,Ke2014}. 
In the present setup, this second thermodynamic observable is the chemical potential $\mu(x)$, 
set equal to zero at the left edge. Rather than plotting $\mu(x)$, in the right panel of 
Fig.~\ref{fig15}, we have preferred to plot the more intuitive mass density $a(x)$.
There we see that the profile for $T_L=1$ deviates substantially from a straight line, suggesting that
the lattice might not be long enough to ensure an asymptotic behavior.

\begin{figure}[ht]
\begin{center}
\includegraphics[width=8 cm,clip]{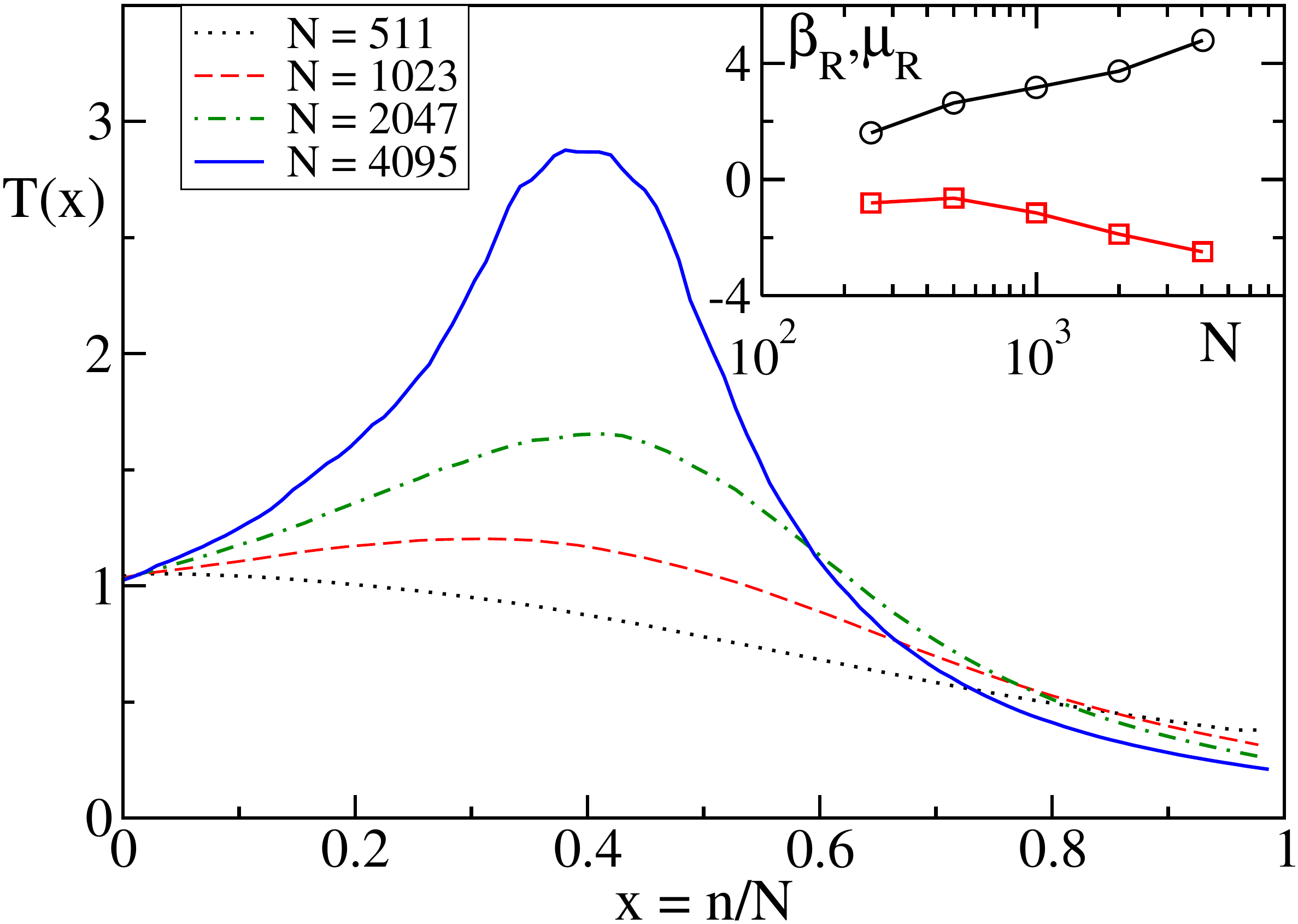}
\end{center}
\caption{ 
Average profiles of the temperature $T(x)$ for $T_L=1$ and different system sizes $N$.
The profile $T(x)$ is computed by means of the microcanonical definition of temperature.
The inset shows the boundary inverse temperature $\beta_R$ (black circles) and chemical potential 
$\mu_R$ (red squares) close to the dissipator side as a function of the system
size $N$. The data refers to the microcanonical definitions of temperature and chemical potential 
computed on the last 10 sites of the chain.
Simulations are performed evolving the system over $10^7$ time units after a transient of 
$4\times 10^7$ units.
For the system size $N=4095$ a further average over 10 independent trajectories has been performed.
}
\label{fig:beta_T1}
\end{figure}

To clarify this point, we have performed simulations for different values of $N$.
The results for $T_L=1$ are reported in Fig.~\ref{fig:beta_T1}, where we plot
the local temperature $T$ as a function of $x$. All profiles start from $T=1$, the value imposed
by the thermostat and, after an intermediate bump, eventually attain very small values.
Since neither the temperature nor the chemical potential are directly imposed by the purely 
dissipating ``thermostat", it is not obvious to predict the asymptotic value of the temperature
(and the chemical potential). The data reported in the inset suggest a sort of logarithmic
growth with $N$, but this is not entirely consistent with the results obtained for $T_L=3$ (see below).

If transport were normal and $N$ were large enough, the various profiles should collapse onto each other, 
but this is far from the case displayed in Fig.~\ref{fig:beta_T1}. The main reason for the lack of convergence 
is the growth of the temperature bump. This is because, upon further increases of $N$, the system spontaneously 
crosses the infinite temperature line and enters the negative-temperature region. For $T_L=3$, this ``transition" 
has already occurred for $N=4095$, as it can be seen in Fig.~\ref{fig15}.
The crossings of the infinite temperature points ($\beta = 0$) at the boundaries of the negative 
temperature region correspond to infinite (negative) values of the chemical potential $\mu$ in such a way that 
the product $\beta \mu$ remains finite at these turning points (data not reported). 

\begin{figure}[ht]
\includegraphics[width=8 cm,clip]{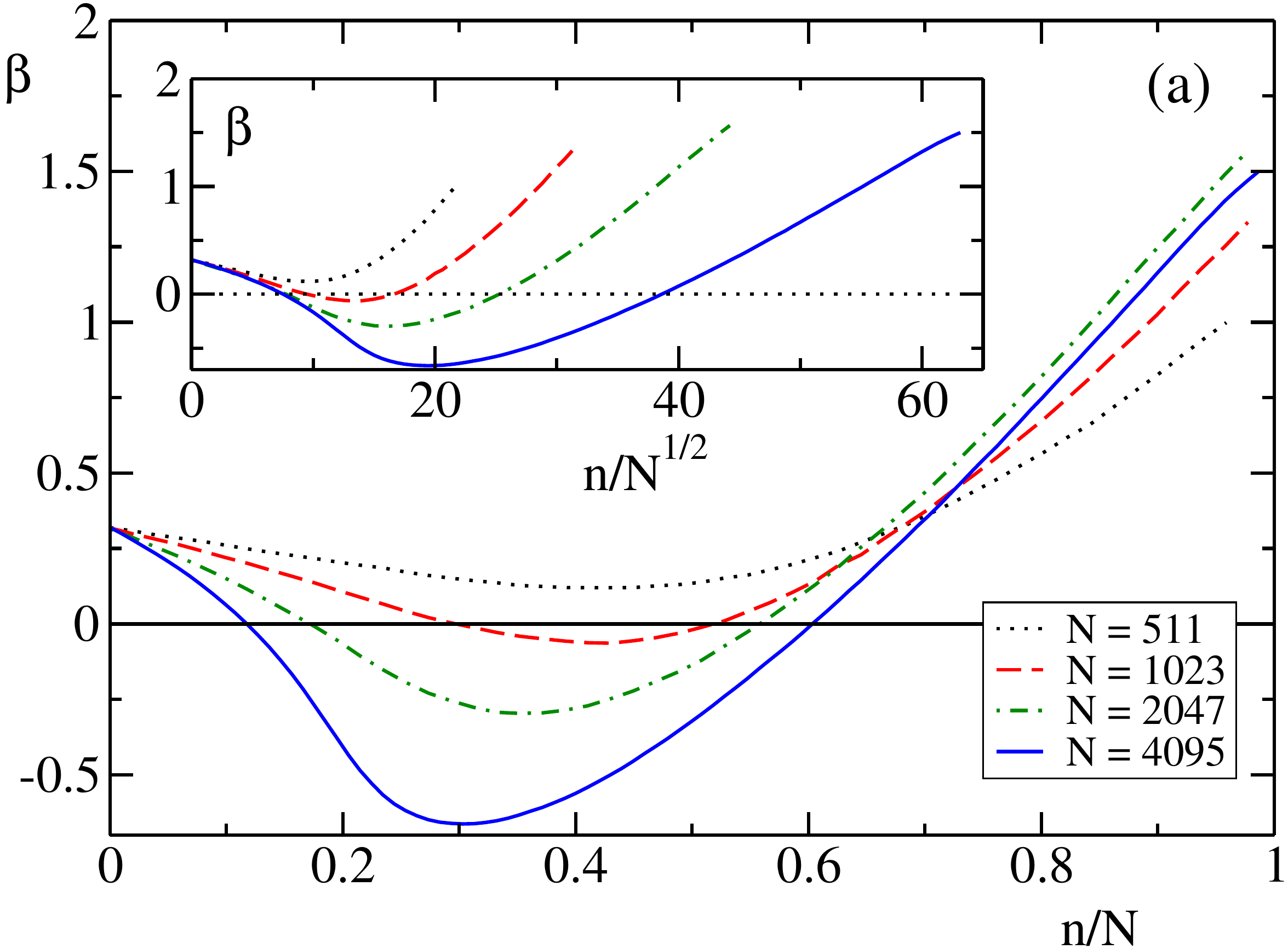}
\includegraphics[width=8 cm,clip]{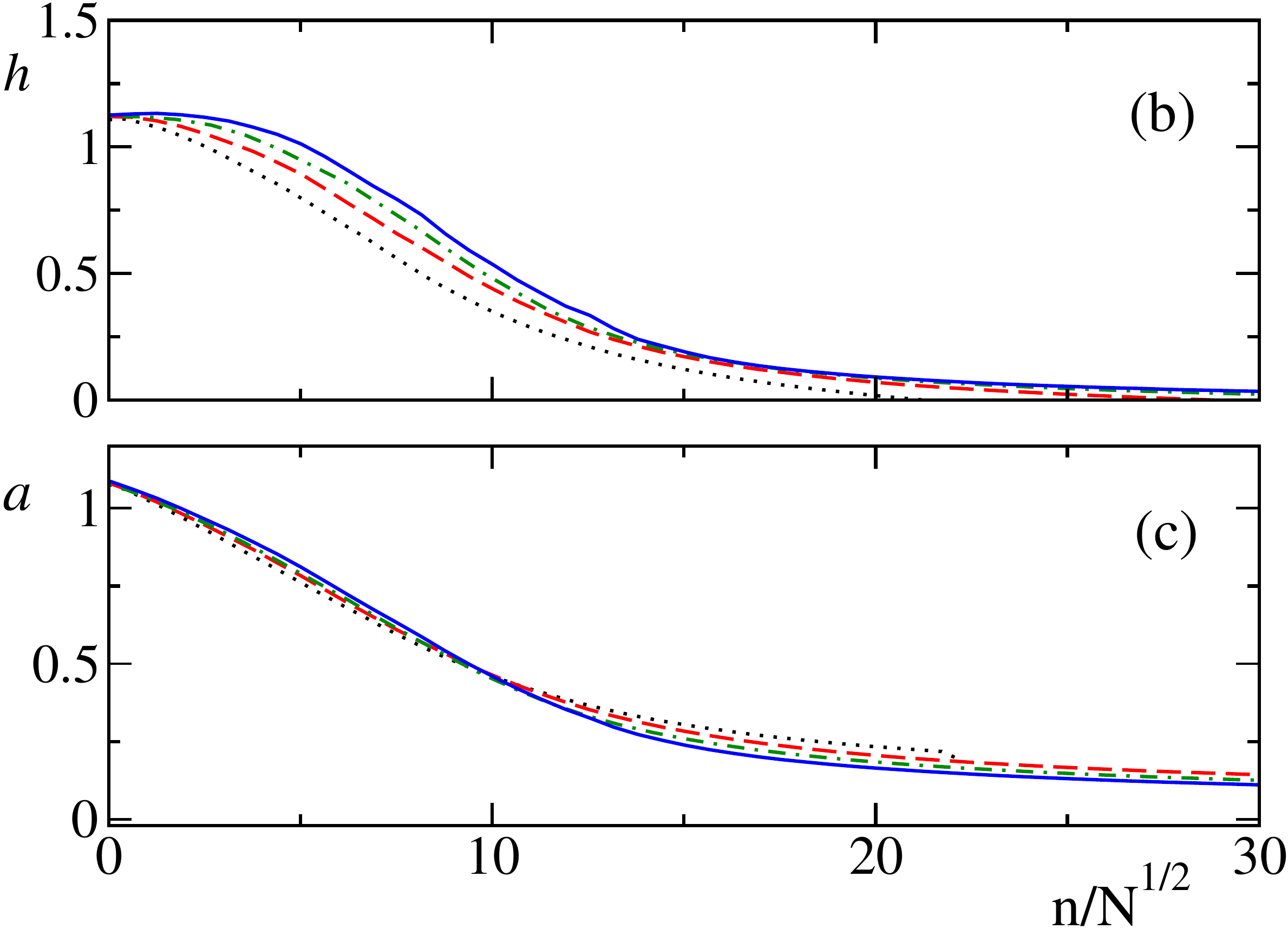}
\caption{Panel (a): average profiles of inverse temperature $\beta$ for $T_L=3$ and different system 
sizes $N$. The profile $\beta$ is computed by means of the microcanonical definition of temperature.
In the inset an alternative scaling by $1/\sqrt{N}$ is proposed.
For the same setup, panels (b) and (c) show the behavior of the energy profile $h$ and the mass profile
$a$, respectively (again scaling the position by $\sqrt{N}$).
Simulations are performed evolving the DNLS chain over $10^7$ time units after a transient 
of $4\times10^7$ units. For the system size $N=4095$, a futher average over ten independent 
trajectories is performed.
}
\label{fig14}
\end{figure}

To our knowledge, this is the first example of negative-temperature states robustly obtained and maintained in
nonequilibrium conditions in a chain coupled with a single reservoir at positive temperature. 
In order to shed light on the thermodynamic limit, we have performed further simulations for different
system sizes. In Fig.~\ref{fig14} we report the results obtained for $T_L=3$ and $N$ ranging from
511 to 4095. In panel (a) we see that the negative-temperature region is already entered for 
$N = 1023$. Furthermore, its extension grows with $N$, suggesting that in the
thermodynamic limit it would cover the entire profile but the edges. 

Since non-extensive stationary profiles have been previously observed
both in a DNLS and a rotor chain (i.e., the XY-model in $d=1$) at zero temperature
and in the presence of chemical potential gradients \cite{Iubini2014}, it is tempting to test
to what extent an anomalous scaling (say $n/\sqrt{N}$) can account for the 
observations. In the inset of Fig.~\ref{fig14}(a), we have rescaled the position along
the lattice by $\sqrt{N}$. For relatively small but increasing values of $n/\sqrt{N}$ we do see
a convergence towards an asymptotic curve, which smoothly crosses the $\beta=0$ line.
This suggests that close to the left edge, positive temperatures extend over a range of order
$\mathcal{O}(\sqrt{N})$, thereby covering a non-extensive fraction of the chain length.
The scaling behavior in the rest of the chain is less clear, but it is possibly a standard extensive
dynamics characterized by a finite temperature on the right edge.
A confirmation of the anomalous scaling in the left part of the chain is obtained by plotting the profiles of
$h$ and $a$ again as a function of $n/\sqrt{N}$ (see the panels (b) and(c) in Fig.~\ref{fig14}, respectively).

\begin{figure}[ht]
 \begin{center}
 \includegraphics[width=9 cm,clip]{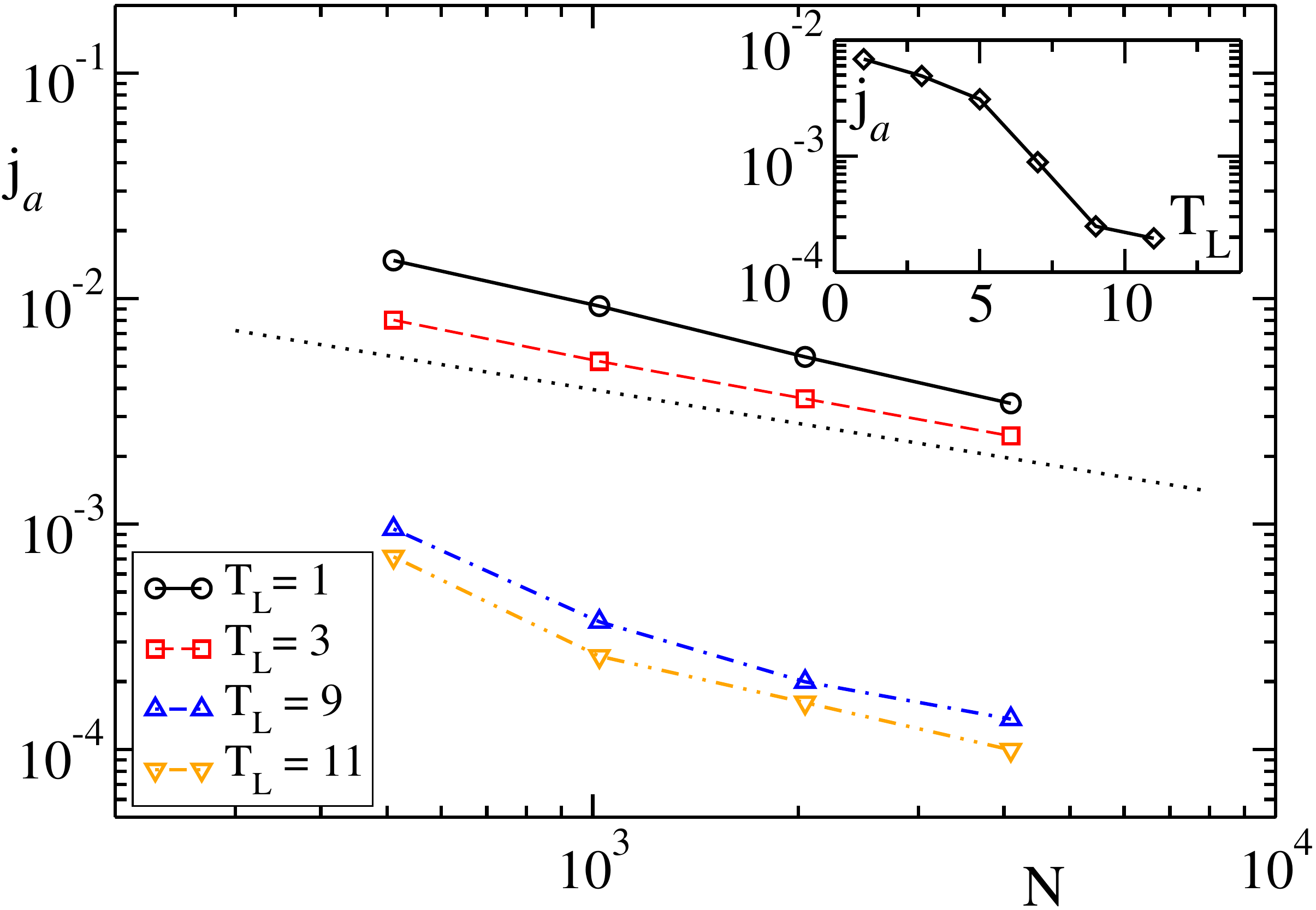}
 \end{center}
 \caption{Average mass flux $j_a$ versus system size $N$ for different 
 reservoir temperatures $T_L$. The black dotted
 line refers to a power-law decay $j_a\sim N^{-1/2}$. 
 The inset shows the dependence of $j_a$ on the reservoir temperature $T_L$ for the system size $N=4095$.
 Simulations are performed evolving the DNLS chain over $10^7$ time units after a transient of 
 $4\times 10^7$ units. For the system size $N=4095$ we have averaged over 10 independent trajectories.
 }
 \label{fig:flux_scal}
\end{figure}

Further information can be extracted from the scaling behavior of the stationary  mass flux $j_a$.
In Fig.~\ref{fig:flux_scal} we report the average value of $j_a$ as a function of the lattice length.
There we see that $j_a$ decreases roughly as $N^{-1/2}$. At a first glance this might be interpreted as
a signature of energy super-diffusion, but it is more likely due to the presence of a pure 
dissipation on the right edge (in analogy to what seen in the XY-model~\cite{Iubini2016}).

\begin{figure}[ht]
\includegraphics[width=8 cm,clip]{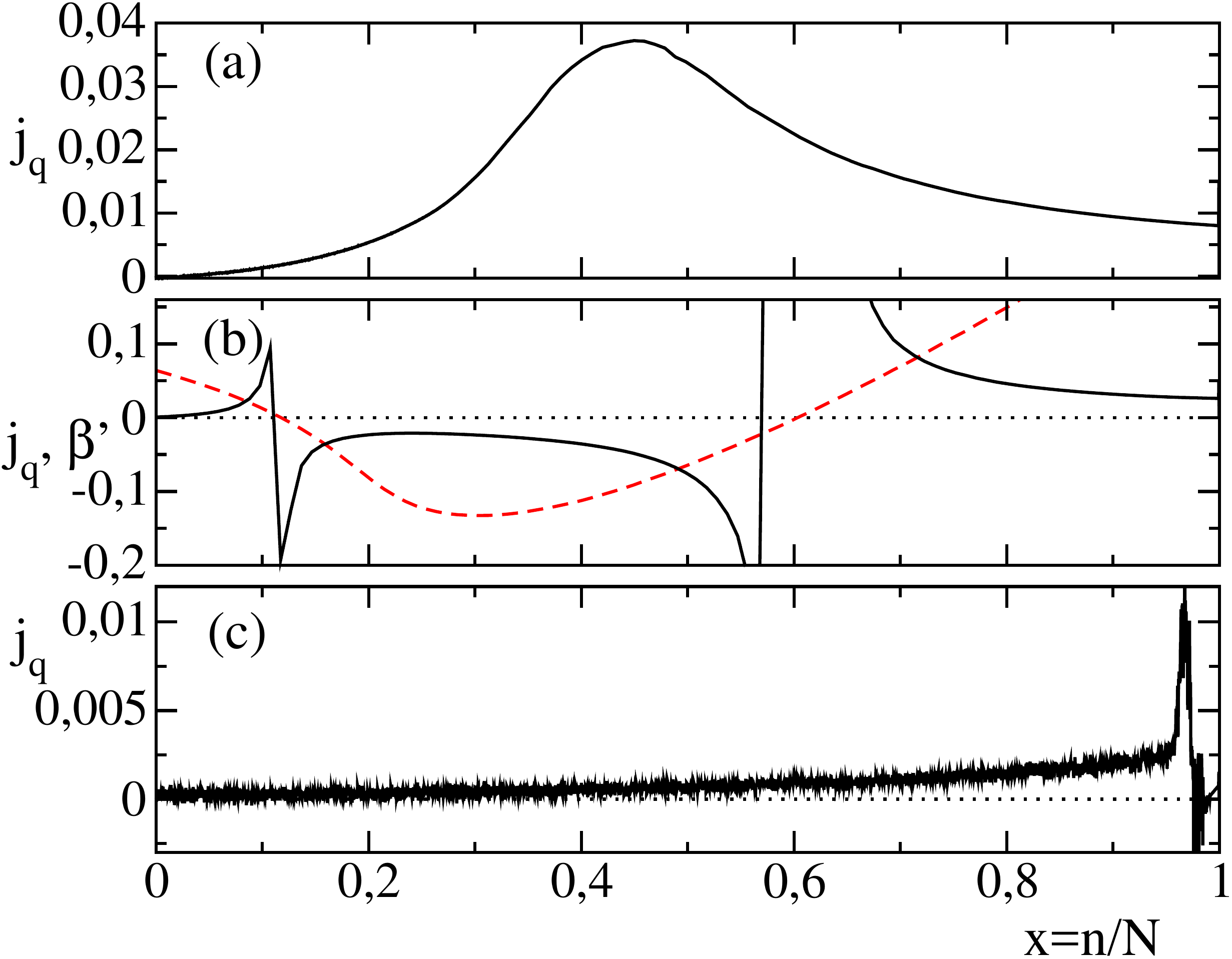}
\includegraphics[width=6.65 cm,clip]{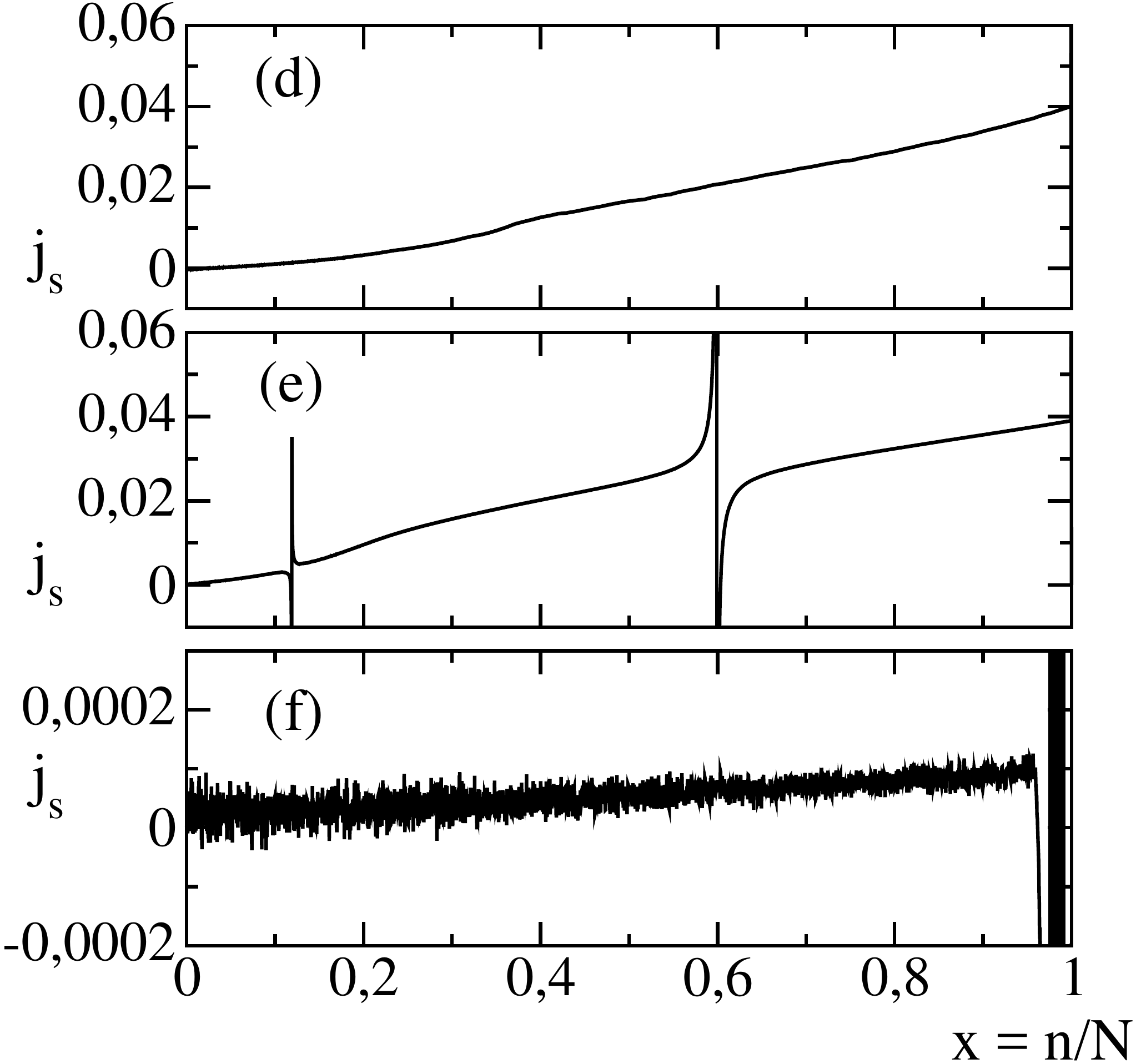}
\caption{Profiles of heat flux (black dashed lines) for $T_L=1$ (panel (a)), $T_L=3$ (panel (b)) 
and $T_L=9$ (panel (c)) in a chain with $N=4095$ lattice sites. For each boundary temperature, we find 
the following values of mass and energy fluxes: panel (a) $j_a=3.5\times 10^{-3}$, 
$j_h=-3.3\times 10^{-4}$; panel (b) $j_a=2.5\times 10^{-3}$, $j_h=3.4\times 10^{-4}$;
panel (c) $j_a=1.2\times 10^{-4}$, $j_h=1.5\times 10^{-4}$. 
The red dashed line in panel (b) refers to the rescaled profile of the inverse temperature 
$\beta'(x)=\beta(x)/5$ measured along the chain (see Fig. \ref{fig15}). 
Panels (d), (e) and (f) show the profiles of entropy flux for the same temperatures: 
$T_L=1$, $T_L=3$ and $T_L=9$, respectively.  
Other simulation details are the same as given in Fig.~\ref{fig15}.
}
\label{fig:heatf}
\end{figure}

In stationary conditions, mass and energy fluxes are  constant along the chain. This is not necessarily true for the heat flux,
as it refers only to the incoherent component of the energy transported across the chain.
More precisely, the heat flux is defined as $j_q(x)=j_h-j_a\mu(x)$ \cite{Iubini2016} 
Since $j_a$ and $j_h$ are constant,
the profile of the heat flux $j_q$ is essentially the same of the $\mu$ profile
(up to a linear transformation). In Fig.~\ref{fig:heatf}(a) we report the heat flux for $T_L=1$.
It is similar to the temperature profile displayed in Fig.~\ref{fig:beta_T1}. It is not a surprise to discover that $j_q$ is 
larger where the temperature is higher. Panel (b) in the same Fig.~\ref{fig:heatf} refers to $T_L=3$. 
A very strange shape
is found: the flux does not only changes sign twice, but exhibits a singular behavior in correspondence 
of the change of sign, as if a sink and and source of heat were present in these two points,
where the chemical potential and the local temperature diverge (see, e.g. the red dashed line in panel (b) 
representing the $\beta$ profile). The scenario looks less awkward if the entropy flux $j_s=j_q/T$ is monitored.
For $T_L=1$, the bump disappears and we are in the presence of a more ``natural" shape (see Fig.~\ref{fig:heatf}(d))
More important is to note that the singularities displayed by $j_q$ for $T_L=3$ are almost removed since
they occur where $T \to \infty$ (we are convinced that the residual peaks are due to a non perfect identification
of the singularities). If one removed the singular points, the profile of the entropy flux $j_s=j_q/T$ has a similar 
shape for the cases $T_L=1$ and $T_L=3$. 
A more detailed analysis of the scenario is, however, necessary in order to provide a solid physical interpretation

\section{High temperature regime: DB dominated transport}
\label{sec3}

Let us now turn our attention to the high-temperature case.
As shown in Fig.~\ref{fig15}(a), for sufficiently large $T_L$ values, the positive-temperature 
region close to the dissipator disappears (this is already true for $T_L=5$) and, at the same
time, the positive-temperature region on the left grows. In other words, negative temperatures
are eventually restricted to a tiny region close to the dissipator side.
This stationary state is induced by the spontaneous formation and destruction of large DBs close to the
dissipator. On average, such process gives rise to locally  steep amplitude profiles that are reminiscent of barriers raised close to the right edge of the chain, see Fig.~\ref{fig15}(b).
As it is well known, DBs are localized nonlinear excitations typical of the DNLS chain. 
Their phenomenology has been widely described in a series of papers where it has been shown 
that they emerge when energy is dissipated from the boundaries of a DNLS 
chain \cite{Livi2006,Franzosi2007,Iubini2013}.  In fact, when pure dissipators
act at both chain boundaries, the final state turns out to be an isolated DB embedded 
in an almost empty background. In view of its localized structure and the fast
rotation, the DB is essentially uncoupled from the rest of the chain and, {\sl a fortiori}, from the dissipators.  
One cannot exclude that a large fluctuation might eventually destroy the DB, but this would
be an extremely rare event.

In the setup considered in this paper, DB formation is observed in spite of one of 
the two dissipators being replaced by a reservoir at finite temperature. 
DBs are spontaneously produced close to the dissipator edge only for sufficiently high values of $T_L$.
Due to its intrinsic nonlinear character, this phenomenon cannot be described in terms of
standard linear-response arguments. In particular the temperature reported in the various figures cannot be
interpreted as the temperature of specific local-equilibrium state: it is at best
the average over the many different macrostates visited during the simulation. Actually, it
is even possible for some of theses macrostates to deviate substantially from equilibrium.
Therefore we limit ourselves to some phenomenological remarks. 
The spontaneous formation of small breathers close to the right edge drastically reduces the dissipation
and contributed to a further concentration of energy through the merging of the DBs into 
fewer larger ones and, eventually to a single DB. Mechanisms of DB-merging have already been 
encountered under different conditions in the DNLS model~\cite{Iubini2013}. 
The onset of a DB essentially decouples the left from the right regions of the chain.
In particular, it \textit{strongly reduces} the energy and mass currents fluxes. 
One can spot DBs simply by looking at the average mass profiles. In Fig.~\ref{fig15}, the presence of a DB is signaled by the sharp peak close to the right edge for both $T_L= 9$ and $T_L=11$.

\begin{figure}[ht]
\begin{center}
\includegraphics[width=10 cm,clip]{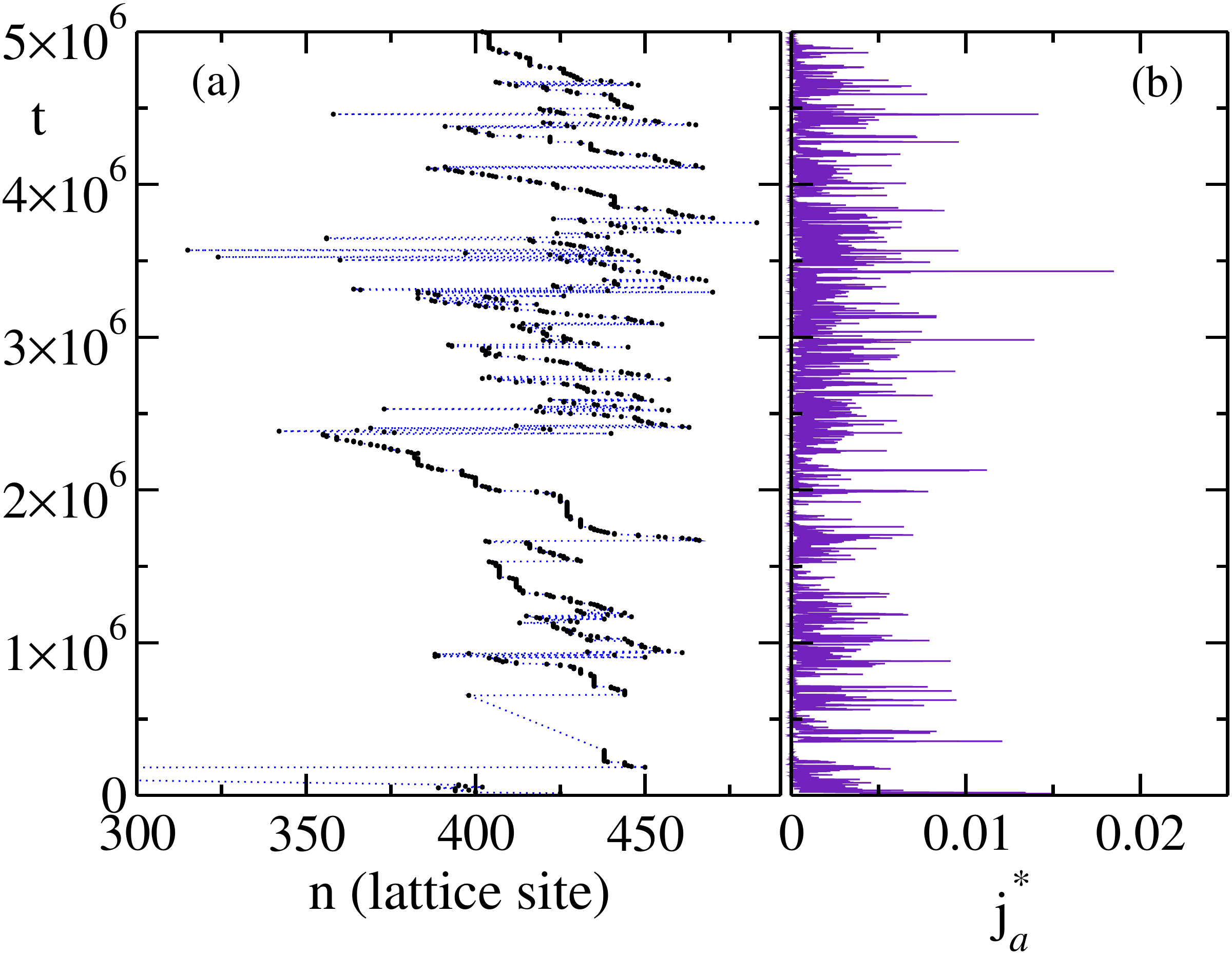}
\end{center}
\caption{(a) Qualitative DB trajectory in a stationary state with $N=511$ and $T_L=10$.
Each point of the curve corresponds to  the position of the maximum average amplitude of the chain in a 
temporal window of $5 \times 10^3$ time units.
(b) Temporal evolution of the outgoing mass flux $j_a^*$   through the dissipator edge
during the same dynamics of panel (a). The flux $j_a^*$ is computed every $20$ time units as the average amount of mass flowing to the  dissipator during such time interval.
Higher  peaks typically correspond to the  breakdown of one or more DBs.
Notice that the boundary mass flux can take only positive values, because the chain interacts 
with a pure dissipator.
}
\label{fig5}
\end{figure}

The region between the reservoir and the DB should, in principle, evolve towards an equilibrium state 
at temperature $T_L$. However, a close look at the  $\beta$-profile in Fig.~\ref{fig15} reveals the presence of
a moderate temperature gradient that is typical of a stationary non-equilibrium state. In fact,
DBs are not only born out of fluctuations, but can also collapse due to local energy fluctuations.
As shown in Fig.~\ref{fig5} (a), once a DB is formed, it tends to propagate towards the heat reservoir located 
at the opposite edge. 
The DB position is tagged by black dots drawn at fixed time intervals over a very long time lapse of
${\mathcal O} (10^6)$, in natural time units of the model. This backward drift 
comes to a ''sudden'' end when a suitable energy fluctuation destroys the DB (see Fig.~\ref{fig5}(a)). 
Afterwards, mass and energy start flowing again towards the dissipator, until a new DB 
is spontaneously formed by a sufficiently large fluctuation close to the dissipator edge 
(the formation of the DB is signaled by the rightmost black dots in Fig.~\ref{fig5}(a))
and the conduction of mass and energy is inhibited again. The DB lifetime is rather stochastic,
thus yielding a highly irregular evolution. The statistical properties of such birth/death process 
are discussed in the following section. 

The statistical process describing the appearance/disappearance of the DB is a complex one. 
On the one hand, we are in the presence of a stationary regime: the mass and energy currents flowing through 
the dissipator are found to be constant, when averaged over time intervals much longer than
the typical DB lifetime. On the other hand, the strong fluctuations in the DB lifetime mean that this regime 
is not steady but it rather corresponds to a sequence of many different macrostates, some of which are likely 
to be far from equilibrium. Altogether, in this phase, the presence of long lasting DBs induces a substantial 
decrease of heat and mass conduction. This is clearly seen in Fig.~\ref{fig:flux_scal}, where $j_a$ is plotted
for different chain lengths. The two set of data corresponding to $T_L=9$ and $11$ are at least one order
of magnitude below those obtained in the low-temperature phase. The sharp crossover that separates the two conduction regimes is neatly highlighted in the inset of Fig.~\ref{fig:flux_scal}, where the stationary mass flux is reported as a function of the reservoir temperature for the size $N=4095$.

The effect of the appearance and disappearance of the DB in the high reservoir temperature regime on the 
transport of heat and mass along the chain is twofold. During the fast dynamics, it produces burts in the 
output fluxes of these quantities as demonstrated in Fig.~\ref{fig5}(b). When the DB is present the boundary flux
to the dissipator decreases, while when the DB disappears, avalanches of heat and mass reach the dissipator. In the 
slow dynamics obtained by averaging over many bursts, the conduction of heat and mass from the 
heat reservoir to the dissipator is hugely reduced with respect to the low temperature regime. We can 
conclude that the most important effect of the intermittent DB in the high temperature regime is to act as 
a thermal wall.

Finally, in Fig.~\ref{fig:heatf} (panels (c) and (f)) we plot the heat and entropy profiles observed in
the high-temperature phase, respectively. The strong fluctuations in the profiles are a consequence of the 
large fluctuations in the DB birth/death events and its motion. It is now necessary to average over much 
longer time scales to obtain sufficiently smooth profiles. It is interesting, however, to observe that the profile 
of $j_s$  in Fig.~\ref{fig:heatf} (f) exhibits an overall shape similar to that observed in the low temperature 
regimes (see Fig.~\ref{fig:heatf} panels (d) and (e)). 
This notwithstanding, there are two main differences  with the low temperature behaviour. First, close to the right 
edge of the dissipator we are now in presence of wild fluctuations of $j_s$, and second, the overall scale of the 
entropy flux profile is heavily reduced.

\section{Statistical analysis}
\label{sec4}

\begin{figure}[ht]
\includegraphics[width=7 cm,clip]{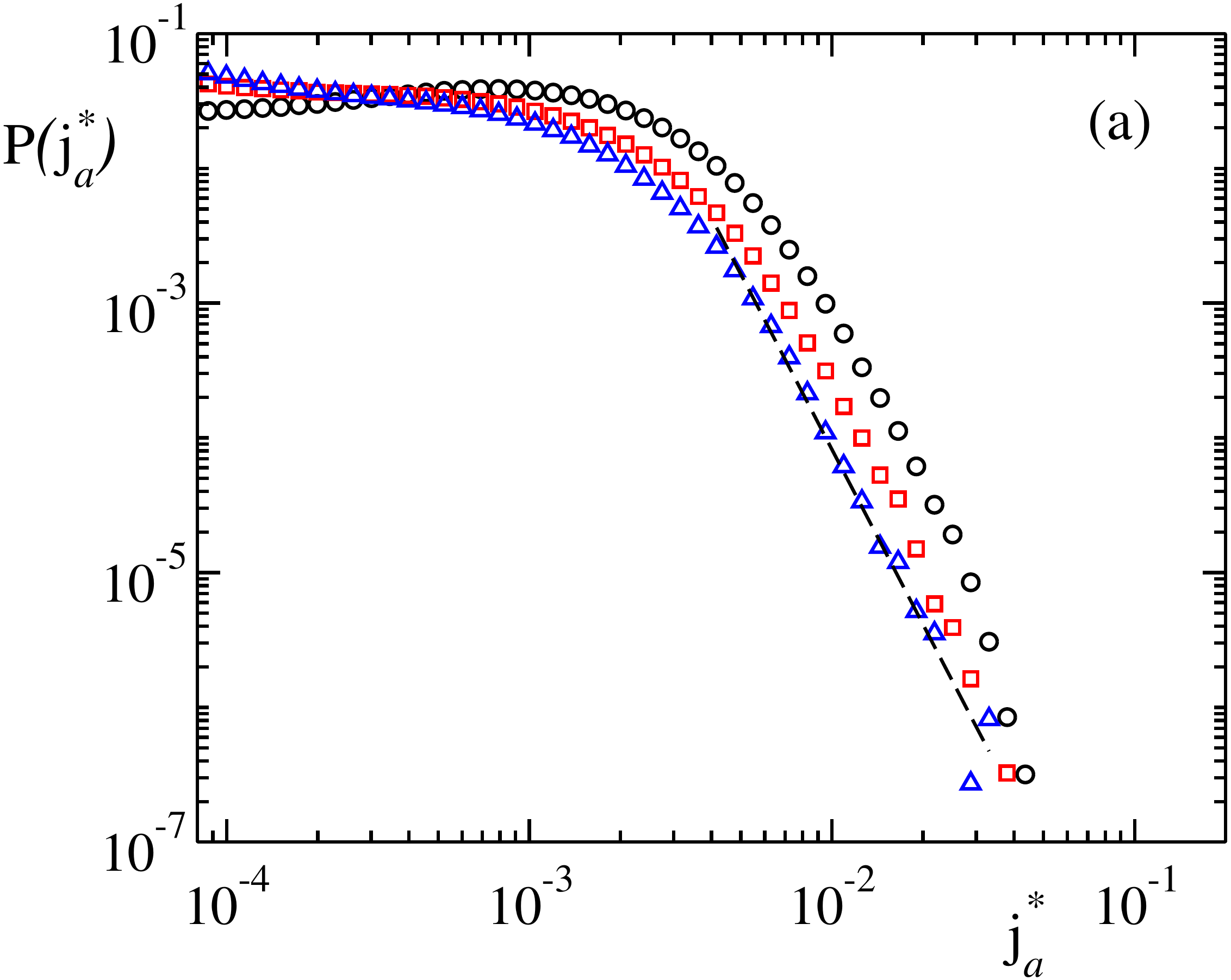}
\includegraphics[width=7 cm,clip]{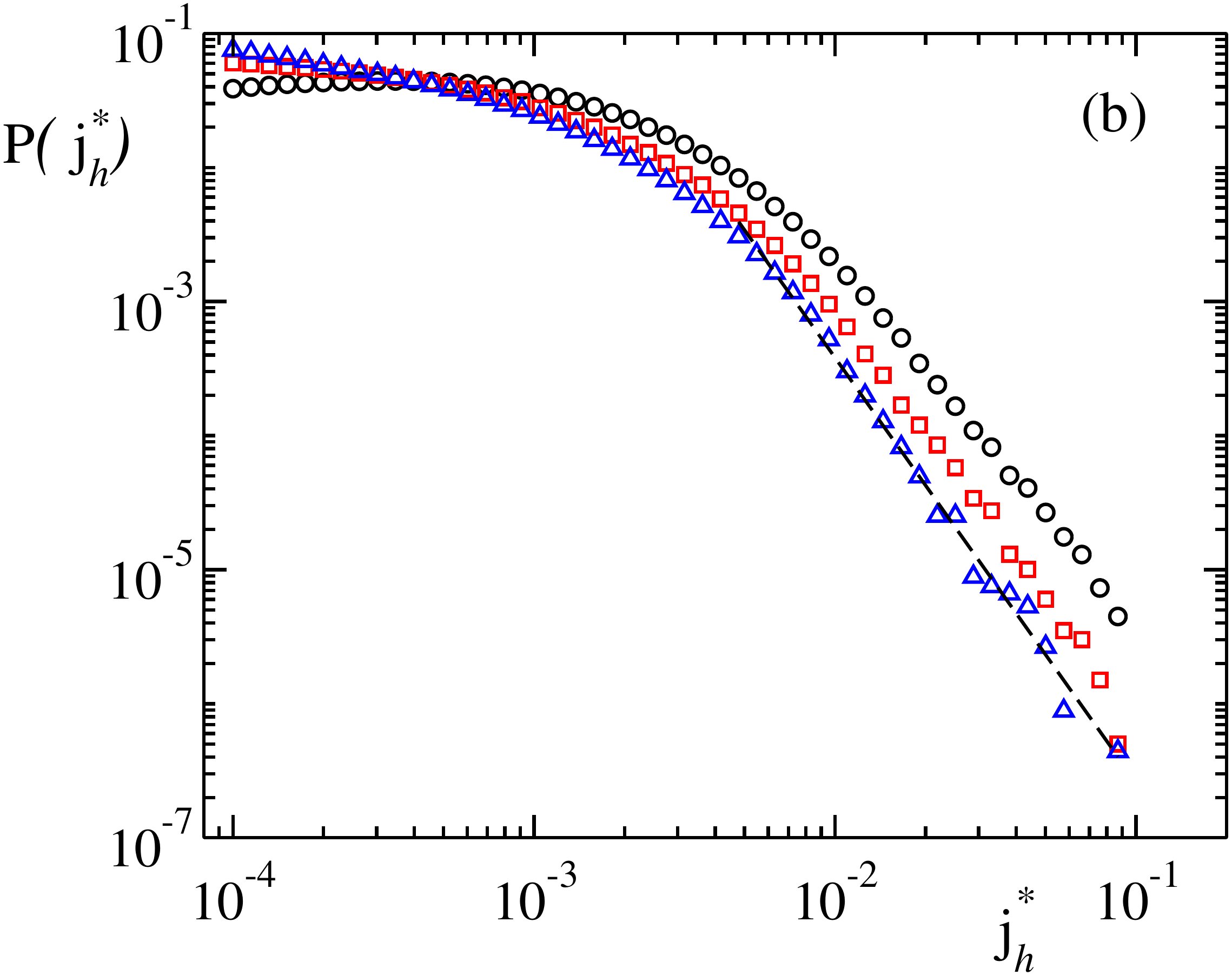}
\caption{Normalized boundary flux distributions  through the dissipator for a stationary state with $T_L=11$. 
Panel (a) shows the mass flux distribution $P\left(j^*_{a}\right)$, while panel (b) refers to the energy  flux  
distribution $P\left(j_h^*\right)$. 
Black circles, red squares and blue triangles refer to system sizes $N=511, 1023$ and $2047$,
respectively. Power-law fits on the largest size $N=2047$ (see black dashed lines) give 
$P(j_a^*)\sim {j_a^*}^{-4.32}$ and $P(j_h^*)\sim {j_h^*}^{-3.17}$.
Boundary fluxes are sampled by evolving the DNLS chain for a total time $t_f$ after a transient of $4\times10^7$ 
temporal units and averaged over time windows of $5$ temporal units.
For the sizes $N=511$ and $N=1023$ we have considered a single trajectory with $t_f= 10^8$.
For the size $N=2047$  the distributions are extracted from 5 independent 
trajectories with $t_f=2\times10^7$.
}
\label{fig:D_Eflux}
\end{figure}

In order to gain information on the high-temperature regime, it is convenient to look at the fluctuations
of the boundary mass flux $j_a^*$ and the boundary energy flux  $j_h^*$  flowing through the dissipator edge.
In Fig.~\ref{fig:D_Eflux} we plot the distribution
of both $j_a^*$ (panel (a))
and $j_h^*$ (panel (b)) for $T_L=11$, for different chain lengths. In both cases, power-law tails almost 
independent of $N$ are clearly visible. This scenario is highly reminiscent of the avalanches occurring
in sandpile models. In fact, one such analogy has been previously invoked in the context of DNLS dynamics
to characterize the atom leakage from dissipative optical lattices \cite{Ng2009}.

We processed the time series of the type reported in Fig.~\ref{fig5} (b) to determine the 
duration $\tau_b$ of the bursts (avalanches) and $\tau_l$ of the ``laminar" periods in between 
consecutive bursts (i.e. the DB life-times). In practice, we have first fixed a flux threshold 
($s=4.25\times 10^{-3}$) to distinguish between burst and laminar periods. 
Furthermore, a series of bursts separated by a time shorter than $dt_0=10^3$ has been treated 
as a single burst. 
This algorithm has been applied to 20 independent realizations of the DNLS dynamics in the 
high-temperature regime. Each realization has been obtained by simulating a lattice with $N=511$ sites, 
$T_L=10$ and for a total integration time $t=5\times 10^6$. In these conditions we have 
recorded nearly $7000$ avalanches.

\begin{figure}[ht]
\includegraphics[width=7 cm,clip]{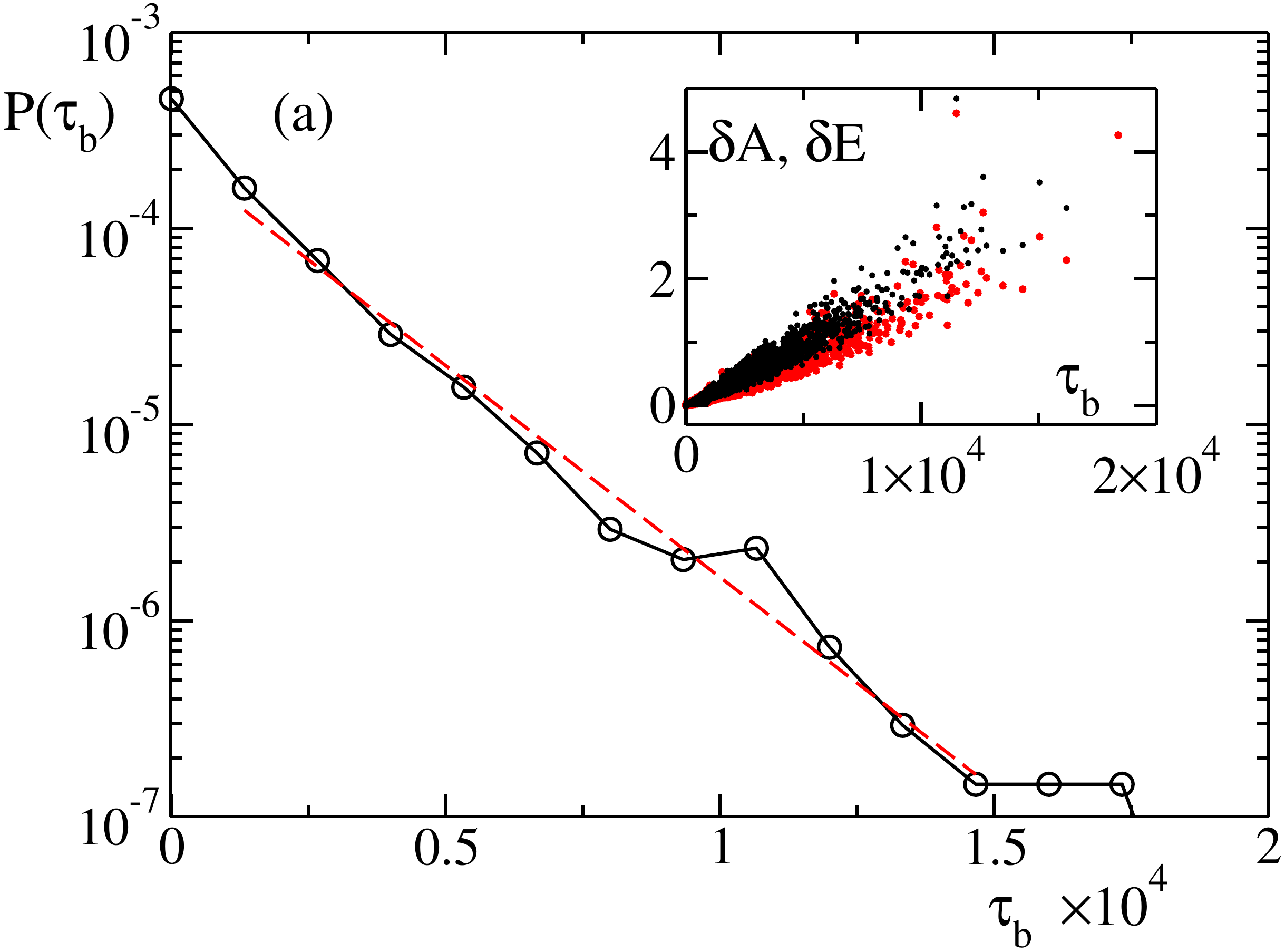}
\includegraphics[width=7 cm,clip]{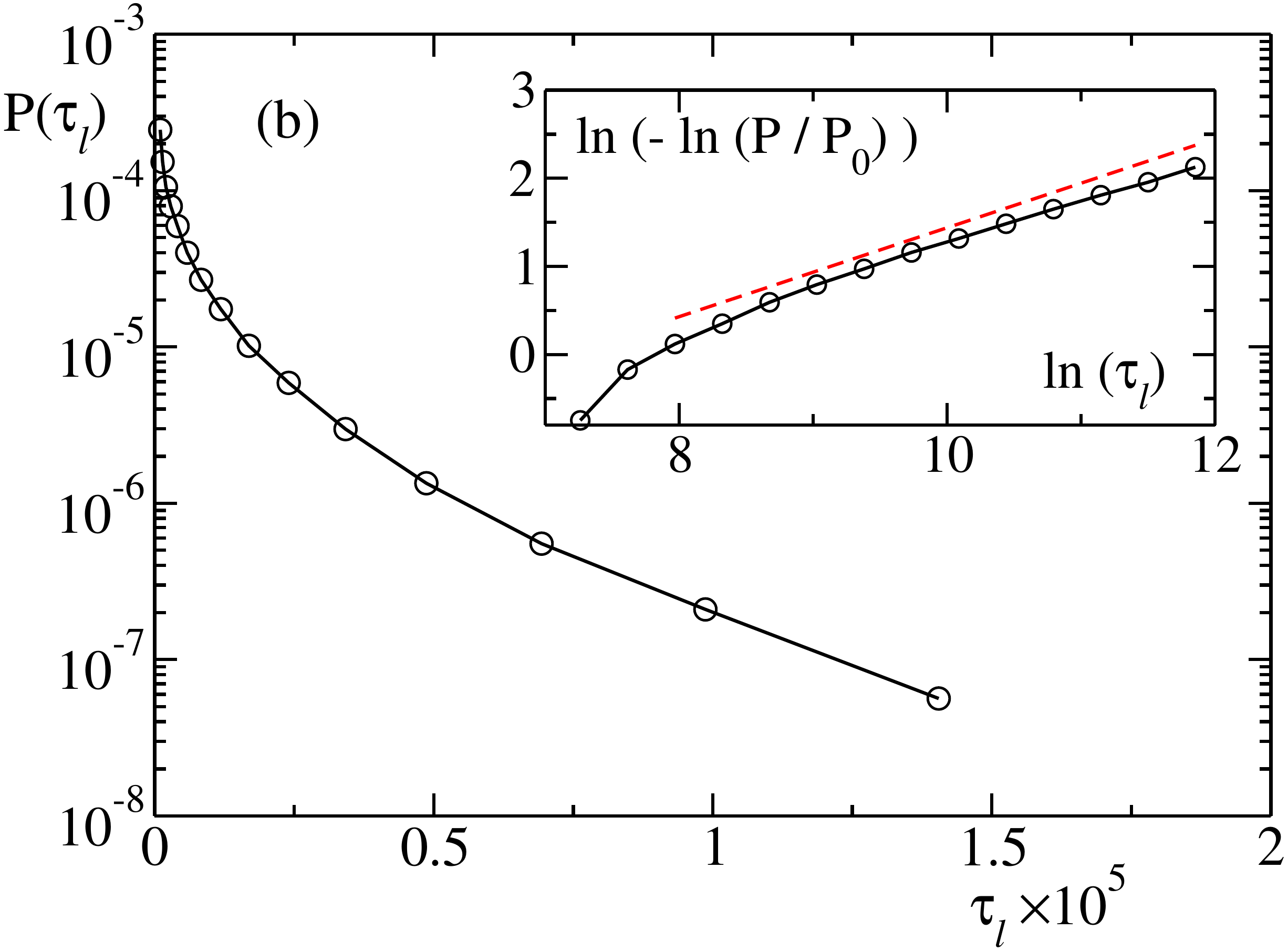}
\caption{(a) Probability distribution of the duration of bursts $\tau_b$. Note the logarithmic scale on the vertical 
axis. From an exponential fit $P\sim \mathrm{e}^{-\gamma \tau_b}$, we find a decay
constant $\gamma=5\times 10^{-4}$ (red dashed line).
The inset shows the relation between the duration $\tau_b$ and the amount of mass $\delta A$ (black dots) 
and energy $\delta E$ (red dots) released to the dissipator.
(b) Probability distribution of DB lifetimes. In the inset we show that this is compatible with a stretched 
exponential law $P=P_0 \mathrm{e}^{-\tau_l^\sigma}$, where  $\sigma\simeq0.5$ (red dashed line) and $P_0$ 
is the maximum value of the distribution. 
}
\label{fig8}
\end{figure}

The probability distribution of the burst duration is plotted in Fig.~\ref{fig8}(a). It follows a
a Poissonian distribution, typical of random uncorrelated events. 
We have also calculated the amount of mass $A$ and energy $E$ associated to each burst, 
integrating mass and energy fluxes during each burst. The results are shown as a scatter plot in 
the inset of  Fig.~\ref{fig8}(a). They display a clear (and unsurprising) correlation between these 
two quantities.

The time interval between two consecutive bursts is characterized by a small mass flux.
Typically, during this period the chain develops a stable DB that inhibits the transfer of 
mass towards the dissipator. Fig.~\ref{fig8} (b) shows the probability distribution of the duration 
of these laminar periods. The distribution displays a stretched-exponential decay 
with a characteristic constant $\sigma=0.5$. Such a scenario is consistent with the results 
obtained in \cite{Piazza2003} for the FPU chain and in \cite{Livi2006} for the DNLS lattice. 
The values of the power $\sigma$ found in these papers are not far from the one that we have obtained 
here (see also \cite{Eleftheriou2005} for similar results on rotor models).

Altogether there is a clear indication that the statistics of the duration of avalanches and walls 
is controlled by substantially different mechanisms. It seems that the death of a 
wall/DB is ruled by rare event statistics \cite{Piazza2003,Eleftheriou2005,Livi2006}, while its birth
appears as a standard activation process emerging when an energy barrier
is eventually overtaken. In fact, when mass starts flowing through the dissipator edge after
the last death of a wall/DB, we have to wait for the spontaneous {\sl activation} of a new 
wall/DB before the mass flux vanishes again. Conversely, the wall/DB is typically
found to persist over much longer time scales and its eventual destruction is determined by a
very rare fluctuation, whose amplitude is expected to be sufficiently large to compete with
the energy that, in the meanwhile, has been collected by the wall/DB during its motion
towards the reservoir. 

\section{Conclusions}
\label{sec5}
We have investigated the behavior of a discrete nonlinear Schr\"odinger equation sandwiched between 
a heat reservoir and a mass/energy dissipator. Two different regimes have been identified upon changing 
the temperature $T_L$ of the heat reservoir, while keeping fixed the properties of the dissipator.
For low $T_L$ and low chemical potential, a smooth $\beta$-profile is observed, which extends 
(in the central part) to negative temperatures, without, however, being accompanied by the formation
of discrete breathers. In the light of the theoretical achievements by Rumpf \cite{Rumpf2004,Rumpf2008,Rumpf2009,Rumpf2007}, we
can say that such regions at negative temperature are certainly
incompatible with the assumption of local thermodynamic equilibrium,
usually invoked in linear response theory of transport processes.
In this sense, despite the smoothness of the profiles and the stationarity
of mass and energy fluxes, such negative-temperature configurations should
be better considered as metastable states. Lacking any numerical evidence
of their final evolution even for moderate chain lengths, we can just
speculate that they could last over astronomically long times. 
Further studies are necessary to clarify whether and how this regime can persist
in the thermodynamic limit. It is nevertheless remarkable to see that negative temperatures are
steadily sustained for moderately long chain lengths. As a second anomaly, we report the slow decrease 
of the mass-flux with the chain length: the hallmark of an unconventional type of transport. 
This feature is, however,
not entirely new; a similar scenario has been previously observed in setups with dissipative
boundary conditions and no fluctuations~\cite{Iubini2014}.

For larger temperatures $T_L$, we observe an intermittent regime characterized by the alternation of
insulating and conducting states, triggered by the appearance/disappearance of discrete breathers.
Note that this regime is rather unusual, since it is generated by increasing the amount of energy
provided by the heat bath, rather than by decreasing the chemical potential, as observed for example 
in the superfluid/Mott insulator transition in Bose-Einstein condensates in optical lattices. 
The intermittent presence of a DB/wall makes the chain to behave as a {\sl rarely leaking pipe}, 
which releases mass droplets at random times, when the DB disappears according to a stretched-exponential
distribution. The resulting fluctuations of the fluxes suggest that the regime is stationary but not
steady, i.e., locally the chain irregularly oscillates among different macroscopic states characterized,
at best, by different values of the thermodynamic variables. A similar scenario is encountered
in the XY chain, when both reservoirs are characterized by a purely dissipation accompanied 
by a deterministic forcing~\cite{Iubini2014}. In such a setup, as discussed in 
Ref.~\cite{Iubini2014}, the
temperature in the middle of the chain fluctuates over macroscopic scales.
Here, however, given the rapidity of changes induced by the DB dynamics, there may be
no well-defined values of the thermodynamic observables. For example during an avalanche, 
it is unlikely that temperature and chemical potentials are well-defined quantities, 
as there may not even be a local equilibrium. This extremely anomalous behavior is likely
to smear out in the thermodynamic limit, since the breather life-time does not probably increase
with the system size; however, it is definitely clear that the associated fluctuations strongly
affect moderately-long DNLS chains.

\vspace{6pt}

\acknowledgments{
This research did not receive any funding.
S.L. acknowledges hospitality of the \textit{Institut Henri Poincar\'e - Centre Emile Borel }during the trimester \textit{Stochastic Dynamics Out of
Equilibrium} where part of this work was elaborated.}
\authorcontributions{S.I. performed the numerical simulations. All Authors contributed to 
the research work and to writing the paper.
}

\conflictsofinterest{The authors declare no conflict of interest.} 

\abbreviations{The following abbreviations are used in this manuscript:\\

\noindent 
\begin{tabular}{@{}ll}
DNLS & Discrete Nonlinear Schr\"odinger\\
DB & Discrete Breather\\
\end{tabular}}

\externalbibliography{yes}
\bibliography{heat,diodo}

\end{document}